# Integration of Clinical, Biological, and Computational Perspectives to Support Cerebral Autoregulatory Informed Clinical Decision Making

## Decomposing Cerebral Autoregulation using Mechanistic Timescales to Support Clinical Decision-Making

J.K.Briggs[1], J.N. Stroh[1], T. D. Bennett[2,3], S. Park[4], D.J. Albers[1,2]

1. Department of Bioengineering, College of Engineering, Design, and Computing, Aurora, CO
2. Section of Informatics and Data Science, Department of Pediatrics, University of Colorado School of Medicine, Aurora, CO
3. Section of Critical Care Medicine, Department of Pediatrics, University of Colorado School of Medicine, Aurora, CO
4. Departments of Neurology and Biomedical Informatics, Columbia University Irving Medical Center, NewYork-Presbyterian Hospital, New York, NY, 10032, USA

## Abstract

Adequate brain perfusion is required for proper brain function and life. Maintaining optimal brain perfusion to avoid secondary brain injury is one of the main concerns of neurocritical care. Cerebral autoregulation is responsible for maintaining optimal brain perfusion despite pressure derangements. Knowledge of cerebral autoregulatory function should be a key factor in clinical decision-making, yet it is often insufficiently and incorrectly applied. Multiple physiologic mechanisms impact cerebral autoregulation, each of which operate on potentially different and incompletely understood timescales confounding conclusions drawn from observations. Because of such complexities, clinical conceptualization of cerebral autoregulation has been distilled into practical indices defined by multimodal neuromonitoring, which removes mechanistic information and limits decision options. The next step towards cerebral autoregulatory-informed clinical decision-making is to quantify cerebral autoregulation mechanistically, which requires decomposing cerebral autoregulation into its fundamental processes and partitioning those processes into the timescales at which each operates. In this review, we scrutinize biologically, clinically, and computationally focused literature to build a timescales-based framework around cerebral autoregulation. We conclude that the myogenic mechanism acts on timescales faster than one second, the endothelial mechanism acts in one to two minutes, the metabolic mechanism will act on different timescales depending on the circumstance but will not be reflected in dynamic CA testing due to timescale constraints, and the neurologic mechanism acts within two seconds. **These timescales have never been synthesized in the literature and lay the foundation for a new framework in which to view and test cerebral autoregulatory function.** This new framework will allow us to quantify mechanistic interactions and directly infer which mechanism(s) are functioning based only on current monitoring equipment, paving the way for a new frontier in cerebral autoregulatory-informed clinical decision-making.

# Introduction

Neurological injuries such as traumatic brain injury and stroke are among the leading causes of death and disability in the United States*(1, 2)*. In 2019 there were over 61,000 traumatic brain injury-related deaths*(3)* and 150,000 stroke-related deaths*(4)*. Adverse outcomes following traumatic brain injury or stroke are often linked to secondary injuries caused by disruptions in cerebral blood flow (CBF)*(5, 6)*. Therefore, a priority in neurocritical care is to maintain optimal CBF at all times*(7, 8)*. However, this is not a trivial task.

CBF is controlled by cerebral vascular resistance, or the radii of all cerebral vessels, and the pressure difference between intracranial pressure (ICP) (or cerebral venous pressure, whichever is larger) and arterial blood pressure (ABP). This pressure gradient is called the cerebral perfusion pressure (CPP). Fluctuations in CPP will result in proportional changes to CBF unless the vascular resistance is also changed. Interestingly, the alteration of vascular resistance to oppose changes in CPP, thereby allowing for steady CBF, occurs naturally by means of a physiologic phenomenon known as cerebral autoregulation (CA)*(2, 9)*. Impairment of CA is common after injury*(2)* and strongly associated with worse functional and global patient outcomes*(10–12)*.

To avoid disruptions in CBF after injury, current traumatic brain injury care guidelines recommend keeping a patient's CPP between 60-70 mmHg*(13, 14)*, or greater than 40 mmHg in children*(15)*. However, retrospective studies suggest that patient outcomes can be improved by targeting CPP to where CA function is optimized, known as CPPopt*(16–23)*. Multiple obstacles inhibit the realization of such a personalizable protocol in most clinical settings. First, estimating the value of CPPopt is often unsuccessful when patients have impaired CA*(24)*. Second, current techniques and metrics used to assess CA can be unsafe for critical care patients and often rely on data that is highly sparse and/or collected invasively. Third, recent studies show that current CA metrics are nonconcordant or have a low correlation with one another*(25–28)* and are strongly affected by noise*(26)*. Fourth, studies disagree on whether different categories of testing techniques, known as dynamic CA (dCA) and static CA (sCA), yield compatible or differing CA quantifications.

We hypothesize that one source of these many barriers to accurate CA quantification is the complexity of CA itself: CA is a physiological process that operates on multiple temporal and spatial scales and includes several pathways. These complexities are not accounted for when computing and interpreting CA-related measures. To ameliorate these complications and better quantify CA, we must have a unified framework or model anchored to an understanding of the physiologic subprocesses and their timescales. This framework requires a close look at the temporal dynamics of CA physiology and the implicit timescale assumptions made in tests and metrics. Such a unified discussion is currently missing in the literature.

A first step toward biological understanding and modeling a complex system such as CA is to decompose it into a collection of fundamental processes and then isolate and partition those processes into the timescales at which each operates*(29)*. To do this, we will assume the underlying physiologic mechanisms are temporally disjointed, and there is no interaction across timescales. This assumption is likely incorrect at some spatio-temporal scales, but it allows us to quantify CA explicitly and permits physiologic interpretation with the price of reducing realism and negating interactions across timescales. Moreover, by analyzing the spatial and temporal scales that lead these assumptions to fail, we can learn the scales at which the mechanisms interact, directing future diagnostic and model development.

A physiologically-informed CA framework also provides clinically helpful knowledge by clarifying which CA sub-mechanism(s) a given metric or test quantifies given the timescale each test/metric assumes. Drawing these associations allows us to address the inconsistencies in current CA tests and metrics, which arise if a quantification incompletely reflects the total CA response and is the first step toward informing treatment to directly improve the functionality of specific CA mechanisms.

This review integrates basic science and clinical literature to start building a temporally-partitioned CA identification framework. We begin by reviewing the current theories about CA mechanisms and the tests and metrics commonly used to quantify their functionality. The metrics and tests analyzed are not exhaustive but include the most widely studied and clinically implemented. Next, we synthesize this information to conclude the timescales associated with each mechanism and refine the interpretation of CA assessments related to the mechanism they may be testing.

## Background
### Physiologic Mechanisms

The myogenic, endothelial, metabolic, and neurologic responses*(16, 25, 30, 31)* are the four primary mechanisms hypothesized to trigger vasoconstriction or dilation as part of CA. Pericytes also play a role in blood flow regulation at the capillary level, but they will not be considered in this review due to their highly localized nature*(32)*. In isolation, none of the four mechanisms fully explain the CA phenomenon*(9, 10)* and the details of their interactions and dynamics are not fully understood.

Identifying which physiologic mechanisms contribute to CA is an active topic of research. This is primarily because CA has been defined in multiple ways. When strictly defined as the vasculature's response to a change in pressure, CA can only include the myogenic mechanism*(2, 34)*. However, traditional CA testing (described later), measures the emergent dynamics of all of these mechanisms. As such, all four mechanisms much be included when discussing CA. Table 1 presents a more comprehensive summary of these mechanisms.

Myogenic

The myogenic mechanism is the only CA mechanism directly triggered by changes in blood pressure. When the transmural pressure increases, stretch-sensitive ion channels in the smooth muscle of the vessel wall open, allowing for calcium influx and vessel contraction*(31)*. While early studies indicated that the smaller arteries were primarily responsible for the myogenic response in the cerebral vasculature*(35, 36)*, later studies showed that larger arteries in the circle of Willis and pial arteries with diameters larger than 200 $\mu$m may also play a role in myogenic regulation*(37, 38)*. Therefore, when the larger arteries (which are upstream of smaller arterioles) are capable of fully compensating for the change in blood pressure, the myogenic mechanism acts as the first-line of defense*(31, 37, 38)* against changes in CPP. This then allows the smaller arterioles to act as "fine-tuning" vessels, carefully adjusting their diameter according to the specific metabolic demands of the tissue.

*Endothelial*

According to the endothelial mechanism theory, when mean ABP (mABP) increases, it causes an increase in cerebral blood flow velocity (CBFV), which increases shear stress on the endothelium. The endothelial mechanism responds to this shear stress by releasing nitric oxide, which triggers vasodilation*(30)*. Conversely, a decrease in shear stress does not directly affect vascular diameter (although if nitric oxide is present in the blood and vasculature, it's reduction in concentration due to decreased shear stress will cause relaxation of the dilation). Independent of mABP, shear stress is also dependent on blood viscosity, which can be altered naturally, such as when blood glucose concentration or hematocrit concentration increases, or by medical intervention, such as mannitol or hypertonic saline administration. This is a helpful example of how some mechanisms are intertwined with CA, but the action of that mechanism does not always indicate CA action (under the definition of CA responding to changes in CPP). For example, the endothelial mechanism triggers vasodilation when CPP increases (e.g. CA), but it will also trigger dilation when blood viscosity increases (not CA). Importantly, the endothelial mechanism can oppose the effects of the myogenic mechanism: myogenically-initiated vasoconstriction increases shear flow within the vessel, to which the endothelial response is vasodilation and vice-versa.

*Metabolic*

Metabolic regulation occurs when the smooth muscle cells comprising the vessel wall and the cerebral parenchymal cells undergo metabolism and release metabolites into the bloodstream. The increased relative abundance of $CO_2$, lactate, pyruvate, and adenosine causes vasoconstriction, while the increased concentration of $O_2$ causes vasodilation. In contrast to other organs, the cerebrovasculature is much more sensitive to $CO_2$ than $O_2$*(31)*. Some standard of care protocols recommend exploiting this mechanism to lower ICP, whereby hyperventilation causes temporarily which triggers vasoconstriction and a lowers of the volume taken up by the vasculature*(39)*.

Metabolic regulation occurs globally and locally depending on the extent and type of metabolite accumulation (which increases naturally as perfusion decreases). In the case of ischemia or hypoventilation, metabolites throughout the vasculature will cause vasodilation globally. Alternatively, when specific brain regions are highly active or local ischemia occurs, the metabolic mechanism will cause vascular diameter changes isolated to the perturbation area.

*Neurologic*

The neurologic mechanism describes the process in which neurotransmitters, including norepinephrine and acetylcholine, trigger vasodilation and vasoconstriction in small to medium-size vessels*(2)*. There is debate over whether the neurologic mechanism fits under the definition of CA*(40–43)*. The argument against the neurologic mechanism fitting under CA states that the neurologic mechanism responds independently of ABP changes*(2)* and, unlike the myogenic or endothelial mechanisms, which are intrinsically regulated (e.g., chemicals are released from the cells composing the vasculature), the neurologic (and metabolic) mechanisms can be extrinsic to the vasculature (e.g. neurotransmitters are released by neurons and metabolites are released by cerebral parenchyma, which are external to the vasculature). Arguments for the neurologic mechanism to fit under CA definitions observe that the effects of CA have been shown to decrease when neurologic inhibitors are introduced*(33, 42, 44)*.

*Autoregulatory Range*

In one of the first comprehensive studies of CA, Lassen*(45)* postulated that intact CA could only function on a set range of ABP values. This range was henceforth referred to as the autoregulatory range and is generally thought to be between 60-160 mmHg*(46)*. Some have more recently argued that the independent/causal variable should be changed from ABP to CPP*(8)*. While Lassen did not differentiate between CA mechanisms, the autoregulatory range is usually referred to in the context of the myogenic mechanism*(2, 31, 48)*. One reason this contextual interpretation is implied is because the metabolic mechanism is *not* limited by the autoregulatory range, making it highly influential during hypo- or hypertensive events*(2, 49–51)*. The autoregulatory range can be shifted or shortened substantially with injury*(52)* and other factors such as metabolites*(53, 54)*, chronic vasospasm*(55)*, or chronic hypertension*(56, 57)*. Outside of this range, CBF will passively reflect changes in pressure, which Jordan et al.*(58)* refers to as "falsely" impaired CA.

*Mechanism response to ABP changes*

As previously mentioned, there is debate concerning whether each of the mechanisms above fits under CA's definition*(2)*. Recall that CA is usually defined as the response of the vasculature to CPP change. Because the myogenic mechanism is the only mechanism directly triggered by pressure changes, some consider it as the sole contributor to CA*(2, 34)*, while others offer different combinations of these four mechanisms the CA mechanisms*(33, 42, 44, 59–62)*.

Figure 1 shows how each of the four CA mechanisms is canonically hypothesized to respond to a rapid change in ABP. Note that the response of mechanisms to an increase in ABP is not always the inverse of the response to a decrease in ABP. This means that it is inaccurate to use results from a test that measures vasculature response to decreasing ABP as an indicator of vasculature response to increasing ABP, or vice-versa. Additionally, note that each mechanism has different lag times to respond to ABP changes. Thus, if the faster mechanism is working properly, and can fully stabilize blood flow, slower mechanisms will not come into play. These timescales are discussed below.

*Impairment*

To understand how the four mechanisms can be impaired, consider the following examples. First, chronic hypertension will encourage plaque development and a decrease in vessel compliance*(31)*, which can cause endothelial mechanism dysfunction or impairment of the metabolic mechanism*(63)* but not the myogenic mechanism. Second, the metabolic, myogenic, and endothelial mechanisms can still function when neurovascular coupling is removed*(64)*, though CA may be altered*(33, 44)*. Third, the impairment of endothelial mechanism action does not decrease the influence of the metabolic mechanism in hypoxia or hypercapnia*(65)*. Fourth, the myogenic mechanism may be augmented during subarachnoid hemorrhage events. This improper over-reactivity of the myogenic mechanism increases risk of cerebral hypoxia*(66)*. Because these mechanisms can be impaired independently and play different roles in the CA response, it is vital that we eventually learn to diagnose the state of the individual mechanisms to optimize clinical decision-making. For example, impairment of the metabolic mechanism is more rare and strongly associated with mortality*(67)*, thus tracking the state of the metabolic mechanism may inform choices for clinical intervention.

## Tests and Metrics

Given the complexity of the intracranial system, it is not currently possible to isolate and test the functionality of these four mechanisms in patients. As such, there is no gold standard test to measure CA functionality. Instead, the dynamics of measurable physiology, such as ABP or ICP, are compared with each other to extract information about the system. Figure 2 shows the relationship between CA metrics and the mechanisms they are trying to capture.

*Testing techniques*

CA testing methods are categorized into either dynamic (dCA) or static (sCA) cerebral autoregulation based on timing and circumstance*(68)*. Because there are many cases in which subjects may have impaired CA indicated by a dCA test, but intact CA indicated by an sCA test*(8, 69, 70)*, many view dCA and sCA as subtypes of the CA physiology*(8, 71–73)* rather than simply CA testing methods. In this discussion, we consider dCA and sCA as CA tests, not physiologic subtypes, and investigate which of the above physiologic mechanisms each test is reflecting.

sCA methods measure the cerebral vasculature's response to a slower, more prolonged change in blood pressure*(62, 68)*. These changes can be drug-induced, but clinicians more often rely on blood pressure changes that occur naturally throughout the day, such as those resulting from periodic posture changes. Unlike dCA assessment, sCA techniques do not give information on latency, which has been used as a marker of CA impairment*(68)*. The mechanics of sCA methods are described in detail below in the "Measurement Indices" section.

Alternativeley, dCA methods measure the cerebral vasculature's response to a rapid change in blood pressure*(62)*. This change is often induced by thigh cuff compression, carotid artery compression, or rapid change in body position*(68, 69)*. In thigh cuff measurements, a practitioner applies significant pressure on the femoral artery for multiple minutes and then quickly releases, causing a negative pressure impulse. Carotid compression*(74, 75)* is similar to thigh cuff, except the blood is cut off at the carotid artery. The carotid compression test is free of influence from reflexes inferior to the cerebral vasculature, such as the baroreflex*(38)*, which may confound the thigh cuff results. In all these techniques, the change in CBF is compared to the induced change in ABP. The extent to which CBF changes based on this pressure impulse and the time for CBF to be restored is then used to quantify dCA. These techniques can be dangerous to patients and rely on negative impulses to quantify CA.

*Measurement Indices*

Based on these sCA and dCA tests, different quantifications, or metrics, have been used to describe CA functionality. The most common sCA metrics include the pressure reactivity index (PRx) and mean flow index (Mx). A dCA test can directly determine CA function without further analysis by considering CA impaired if CBF passively follows ABP or intact if the restoration of CBF occurs before ABP is restored. An additonal dCA metric is the autoregulatory index (ARI), which quantifies CA impairment on a scale from 0-9. We will focus on these metrics in this review. We present a summary of other CA metrics in Table 2.

PRx is calculated by first averaging ABP and ICP over some window length (<60 seconds but exact length varies), then calculating the Pearson correlation coefficient between these averaged signals*(48, 76)*. For the PRx computation to result in CA quantification, it is assumed that impaired CA linearizes the relationship between ABP and ICP because CA can no longer appreciably influence CBF. For this to be true, the compensatory reserve must be depleted. The compensatory reserve refers to movement of cerebral spinal fluid (CSF) out of the intracranial space to decrease the total volume inside the skull and decrease ICP. At a high level, when CSF can no longer drain out of the intracranial space, the compensatory reserve is depleted*(77)*, which must be the case for CA to be the only mechanism driving the translation of ABP to ICP. Alternatively, if one does not make any assumptions about CSF, PRx values will reflect both CA functionality and the compensatory reserve. Within these constraints, PRx values less than zero indicate well-functioning CA, while positive values indicate CA impairment.

Mean Velocity index (Mx) is another metric used to quantify CA and is thought to measure CA more directly than PRx. Similar to PRx, Mx is calculated as the Pearson correlation coefficient between CBFV and either ABP or CPP*(78)*. ABP is used more often to compute Mx when ICP, a more invasive measurement, is not available*(26, 79)*, however as the flow is driven by the pressure gradient rather than ABP alone, it is more accurate to use CPP in the calculation. There is no consensus whether the use of ABP or CPP results in significantly different Mx values*(80, 81)*. There are many other correlation-based indices like PRx or Mx that use slightly different measurements with the same basic assumptions.

In most cases, when plotted against CPP values, these correlation based indices (PRx, Mx, or some other variation) make a 'U' shaped curve, leading to the theory that there is an optimal CPP ($CPP_{opt}$) at which PRx is minimized*(82)*. Retrospective studies indicate that keeping patients at $CPP_{opt}$ is associated with better outcomes*(23, 82)*. However, CPPopt may not be successfully identified for patients with impaired CA (high PRx), absent ABP slow waves, who have not received neuromuscular blockade medications, or in the setting of interventions such as decompressive craniectomy or high-dose vasoactive medications*(24)*.

The last metric that we will discuss is the autoregulatory index (ARI), which is a common dCA technique used to quantify CA functionality. ARI was developed by Tiecks et al.*(68)* and is a model-based assessment index based on the rate of change of ABP:CBFV (or equivalently the change in total cerebrovascular resistance, assuming ICP is fixed). This index ranges from 0-9, indicating absent and hyperactive CA, respectively. Each ARI has set parameter values for damping factor, dynamic gain, latency period, and dynamic rate of autoregulation used to map mABP to CBF. When ARI is used to quantify CA in practice, ABP is transformed into ten theoretical $\widehat{CBF}_{ARI}$ values via ARI calculation then these $\widehat{CBF}_{ARI}$ estimates are compared to the measured CBF value. The ARI value that minimizes the difference between this translated CBF and $\widehat{CBF}_{ARI}$ represents the subject's CA functionality*(26, 28, 83)*.

One obstacle preventing broad clinical application of many of these metrics is the availability and reliability of data. For example, CBFV data used to compute Mx and other metrics is obtained by transcranial doppler ultrasound (TCD), which is only available at hospitals with the necessary equipment and expertise. In addition, TCD fails on 5-10% of patients (particularly elderly women) due to lack of temporal bone windows*(84)*. While TCD benefits from being noninvasive, its collection is not automated and requires a technician to take ultrasound measurements. Therefore, even at hospitals with TCD expertise, it is typical for TCD data to be collected once each day for approximately 10 minutes*(85)*, which is far from enough data to compute any of the aforementioned metrics. Additionally, some metrics first convert CBFV to CBF*(68)* through the linear equation $CBFv = \alpha * CBF$, where $\alpha$ is the area of the artery that the TCD device is measuring at that specific depth. It is assumed that $\alpha$ is constant throughout the test, which is not true in situations of changing pCO2*(86)*. This assumption also ignores the vessel taper and the vessel's distensibility, which causes CBFV to slow down while the vessel is being stretched during the peak pressure and speed up while the vessel is recoiling.

The interrelationships between these tests and metrics are not obvious, but the lack of correlation between metrics is clearly established*(25, 52)*. We propose that this lack of correlation can be clarified by investigating the timescale assumptions underlying these metrics and the timescales of physiologic mechanisms themselves. These metrics' timescales and the physiologic mechanisms they are meant to capture are not always explained explicitly but can be found in the available literature.

## Physiological Mechanism Timescales

To hypothesize the timescales behind CA, we identified experimental papers that most successfully isolate an individual mechanism to minimize results confounding mechanism interactions. A synthesis of this section is presented in table 1 and depicted in figure 2. Table 1 presents a summary of the four proposed mechanisms, the hypothesized timescales on which they act, that mechanism's response to change in ABP (assuming that all other physiology is static), and the proposed interaction between the given mechanism and the other three.

### Myogenic Timescale

Symon et al.*(49)* studied the myogenic mechanism in baboons by increasing pressure into the middle cerebral artery while controlling atrial PCO2 and CBF and observed vasodilation as soon as 0.8-1 sec after pressure increase. Similarly, Garcia-Roldan & Bevan*(87)* studied the myogenic mechanism in pial rabbit artery segments by increasing transmural pressure and conserving flow. They observed an immediate 50% increase in diameter.

Endothelial Timescale

Izzard et al.*(88)* showed that when blood flow is altered in rat vasculature, there is a $55\pm18$ second latency before vasodilation begins and 60-120 second latency until maximum dilation is reached. In humans, Black et al.*(89)* used arm cuff deflation to measure flow-mediated dilation and found the time to maximal dilation in subjects younger than 30 years old was ~60 second, and subjects over 50 years old was ~80 seconds. To explain the slower response of the endothelial mechanism compared to the myogenic mechanism, Tong et al.*(90)* postulated that the vessel only responds to increased shear stress after nitric oxide concentration reaches some threshold value, which requires a prolonged latency period.

Metabolic Timescale

The relationship between the metabolic mechanism and CA testing is more complicated than for the myogenic and endothelial mechanisms because the metabolic mechanism will only trigger vasodilation when there are high concentrations of metabolites*(31)*, which occurs globally only in extreme ischemia or when cerebral metabolic rate is altered. The only metabolite shown to influence vascular diameter following CPP changes *within* the autoregulatory range is adenosine*(56)*, a byproduct of ATP dephosphorylation. Its production begins about 5 seconds after a decrease in CBF or increased cerebral metabolic rate*(91)* and increases 3 fold after 1 minute of cerebral artery occlusion*(92)*. However, adenosine-induced vasodilation is concentration-dependent and requires a significant decrease in CBF to have a noticeable impact on vascular diameter*(91–95)*.

For other anaerobic metabolites, such as $H^+$, to accumulate in high enough concentrations to trigger noticeable vasodilation (as a direct result of CPP change), the vasculature must transition to anaerobic metabolism. This transition will only occur when CBF is extremely diminished. Specifically, anaerobic glycolysis and tissue acidosis do not begin until CBF reaches half its resting value ($\sim 25 \frac{ml}{100g/min}$)*(89)* or if cerebral tissue partial pressure of oxygen (PO2) drops below 30 mmHg*(9)*. Even when CBF reaches diminutive values, it must be decreased for a long enough time for the cerebral parenchyma to exhaust its supply of O2 and aerobic nucleotides including NADH. The exact time for this exhaustion to occur is not known. However, oxygen supply in the brain lasts ~30 seconds following total cerebral ischemia *(96)*, so it would take multiple minutes for a mere decrease in blood flow to cause a complete transition to anaerobic glycolysis. Therefore, in traditional CA testing, in which CPP is altered, the impact of the metabolic mechanism would only be seen if CPP is decreased significantly, and no other CA mechanism is functional enough to compensate for the decrease. The relative abundance of these metabolites can be determined using microdialysis or estimated with pCO2 and pO2.

Typical dCA testing will not quantify the functionality of the metabolic mechanism without causing harm to the patient, therefore, we analyzed other clinical interventions to derive its timescales. Auer*(97)* showed a 20 second latency until vasodilation after $CO_2$ inhalation in cats. In humans, Settakis et al*(98)*. showed that vasodilation begins ~10 seconds after breath-holding and reaches its maximum within in 30 seconds. Also in humans, Severinghaus and Lassen*(99)* and Settakis et al*(98)* showed vasoconstriction will begin ~20 seconds following hyperventilation*(100)* and reaches a maximum value within ~30 seconds. Hyperventilation not only changes CBF directly but also influences the timescales associated with other CA mechanisms. For example, Newell et al.*(101)* showed that moderate hyperventilation in severe

TBI patients significantly increased the latency of CA response and rate of vessel dilation after a thigh cuff test.

A final timescale involving the metabolic mechanism that can be observed in dCA testing is that of reactive hyperemia. When CBF is blocked for $x$ seconds and then returned, the now hypoxic environment will trigger vasodilation for an amount of time proportional to ischemia after the flow is returned*(102)*. Gourley and Heistad showed that as the length of ischemia increased from 5 to 30 seconds, the length of hyperemia increased from ∼19 seconds to ∼51 seconds*(103)*. Reactive hyperemia is thought to occur in order to refill the tissue oxygenation and other nutrient deficit that occurred during ischemia*(9)*. This phenomenon would not be directly considered CA because it occurs after CBF has been restored rather than serving a defensive mechanism against disruption in CBF due to changes in CPP, however the absence of a reactive hyperemic response is thought to be associated with impaired autoregulation *(102)* and its functionality contributes to the calculation of the transient hyperemic response ratio, a dCA measurement*(75)*.

[Neurologic Timescale](#)

We were unable to find a study that quantified the timescales of the neurologic mechanism in response to CA testing, partially because the neurologic mechanism may not fit under the CA definition of changes in CBF due to changing blood pressure. Instead, we analyzed studies that quantify the timescale of the neurologic mechanism in other settings. In Rosengarten et al.*(104)*, subjects sat still in a quiet room and were exposed to flashing light, which increased neural activity. Blood pressure did not change significantly; therefore, results were attributed to the neurologic mechanism. The average time to vasodilation was .29 seconds and maximum dilation was reached ~7 seconds after light exposure. Similar results were shown in Ances et al.*(105)*, which found the onset of vasodilation was around 1-2 seconds after electric forepaw stimulation in mice. Masamoto et al.*(106)* tested local CBF changes after hindlimb stimulation in rats and found CBF began to increase ~0.4s after stimulation and reached maximum values between 2.9-4.3 seconds*(106)* while there was no significant change in blood pressure. These studies indicate that the neurologic mechanism can trigger vasodilation in less than 1 sec. However, in most neurovascular studies, including those above, it is not possible to determine whether these results are directly caused by neurologic signaling or due to an increase in cerebral metabolism, which causes vascular diameter change. Therefore, these results may reflect timescales associated with the metabolic rather than direct neurologic signaling. Additionally, these results only fit under a broader definition of CA that describes the vasculature's adjustment to fit the metabolic needs of the cerebrum, different than traditional sCA or dCA tests that focus on changes in blood pressure.

Other timescales associated with the neurologic mechanism and CA testing include those of certain natural periodic waves in CBF and ICP which are attributed to neurologic regulation and can confound CA testing. Meyer waves, or oscillations in ABP with 10 second periods, have been attributed to sympathetic nervous activation*(107)*. Lundberg (or B) waves, defined as periodic fluctuations in intracranial pressure (ICP) occurring with periods of 30–120 sec, are hypothesized to be evidence of neural pacemakers that may increase the cerebral metabolic rate of oxygen and global CBF*(108, 109)*. Weersink et al. showed that methods for finding the CPP that optimizes CA functionality may fail for patients without these natural slow waves*(24)*. Therefore, while the neurologic mechanism may not be directly measured during CA testing, nor directly included in the pressure-specific definition of CA, its action is highly influential in CA test results.

## Long-term physiologic changes that impact cerebral autoregulation

All the mechanistic timescales discussed above can be considered short-term, as they respond to stimuli with 1.5 minutes. Mechanisms or physiologic changes that act on medium (minutes to hours), long (weeks), and very long-term (years to decades) timescales also impact CA. Medium-term mechanisms include changes in blood viscosity, which influence the action of the endothelial mechanism, or changes in ICP, which alter the autoregulatory range. Drugs, phlebotomy, nutrition, hydration, and many other factors can cause medium-term blood viscosity changes. Mannitol, a diuretic drug that lowers blood viscosity, acts 1-20 minutes after administration, with peak effect seen around 20-60 minutes*(110–113)*. In healthy subjects, ingestion of glucose causes blood viscosity changes within 10 minutes *(114)*. Alternatively, pathologies including edema or disruptions in CSF reabsorption can cause increased ICP which alters the autoregulatory range. Elevations in ICP can be transient (such as during coughing or sneezing), or sustained, lasting from 5 min to 6 or more hours*(105)*. Increased ICP is a significant concern for clinicians. Long-term mechanisms involve changes in the myogenic mechanism's homeostatic baseline pressure. For example, if ABP remains elevated for multiple weeks, the vasculature will return to the original radius and the myogenic mechanism will establish this new pressure as baseline*(9)*. Very long-term mechanisms include vascular remodeling, shifting of the autoregulatory range, and impairment of the endothelial mechanism. These are caused by chronic shear stress, hypertension, and vasospasm*(55, 56)*.

## Temporal assumptions and conclusions from CA tests and Metrics

In addition to CA physiological mechanisms, it is important to analyze the timescales associated with CA tests and metrics. As we've discussed, there is poor correlation between CA metrics*(25, 26, 47)* and disagreement over whether dCA and sCA tests and their derived metrics are measuring the same aspects of the CA phenomenon*(68, 70)*. Knowing the timescales underlying these quantifications may shed light on these apparent discrepancies. The results of this analysis are presented in Table 2.

### dCA tests

Dynamic cerebral autoregulation (dCA) testing refers to the set of techniques that measure the action of CA in response to rapid, transient changes in blood pressure. We present timescale results from four papers that used three different types of dCA tests to measure the timescale of CA response to a dCA test. O'Leary et al*(116)* tested CBF in response to a rapid head-up tilt and concluded that there was a ~2 sec delay before the onset of CA. Using the thigh cuff method, Zhang et al.*(117)* and Aaslid et al.,*(53)* concluded that the onset of CA was near instantaneous (< 1 sec) and caused maximum CBF within 5-8 seconds after cuff release. Using carotid compression, Smielewski et al.*(75)* showed CA responded with a 1-2 second latency time and caused maximum CBF within ~7 seconds of compression release. These studies indicate that the physiologic mechanisms dCA is testing have a 1-2 second latency and stabilize large variations in ~5-8 sec.

### sCA tests

Static cerebral autoregulation testing refers to the techniques that measure CA action during slow, prolonged changes in blood pressure. We present three studies that directly adress the timescales associated with sCA tests. Zhang et al. *(117)* used transfer function analysis to assess the coherence between resting CBF and mABP in 10 healthy patients, and concluded that CA acts most effectively on oscillations with periods longer than 14 seconds, and that CA functionality decreases as periods become shorter. Because mABP was used, this study does not analyze frequencies faster than one heartbeat. Similar results were shown in Hamner et al.*(118)* and Fraser et al. *(119)*. Both studies concluded that CA acts on changes to ABP that last longer than 30 seconds. Hamner et al. also showed that the high pass filter-like CA frequency response was independent of the amplitude of pressure change. Therefore, the difference between dCA and sCA timescales cannot be explained by the amplitude of pressure change. These studies indicate that sCA tests reflect the state of slower mechanisms, acting on timescales longer than 14-30 seconds.

## CA metrics

Among the CA metrics we have discussed, the correlation coefficient-based metrics such as PRx and Mx are the only metrics with implicit assumptions about timescales. These metrics are calculated by averaging over a window (usually 5-10 seconds wide*(48, 120)*) then calculating the correlation coefficient over another moving window (between 3-5 min*(48, 107, 121)*). The widths of these windows varies between studies and the most accurate widths are not currently known. In a retrospective study using data from 131 traumatic brain injury patients, Howells et al.*(107)* showed that patients whose PRx value was calculated using ICP and ABP frequency bands from 15 to 55 seconds (extracted using a bandpass filter) had the best predictive ability of outcome as measured by the Glasgow Outcome Scale. The exact range of the averaging window is incredibly important in the resulting CA quantification because, via sampling theory, the metric can only assess the physiological effects of CA mechanisms occurring at timescales *longer* the averaging window. Thus, correlation-based metrics have an implicit assumption that CA is quasi-stationary on timescales faster than the chosen averaging window.

## Integrating Perspectives into a Timescales-Based Framework of Cerebral Autoregulation

Now that we have laid out the timescales and assumptions underlying the CA tests, metrics, and mechanisms, we can begin to explore some of the discrepancies between CA tests and draw associations between the CA quantifications and the mechanisms themselves.

## Informing dCA and sCA impairment

It is clear that dCA and sCA tests have dramatically different timescales, with dCA tests reflecting timescales faster than 8 seconds and sCA tests concluding timescales slower than 15 seconds. This indicates that these two families of tests are measuring different aspects of the CA phenomenon, which is consistent with studies that report CA impairment from dCA tests and not sCA test*(8)*. This also informs the lack of correlation between ARI and sCA tests*(10)*; as ARI was built using a dCA test, whereas Mx assumes timescales closer to that of sCA tests.

dCA tests consistently conclude a latency of <3 seconds with peak CBF reached < 7 seconds after a stimulus. These timescales correspond with the reported timescales of myogenic or neurologic mechanisms. Because the neurologic mechanism occurs independently of pressure change, dCA tests likely measure only the myogenic mechanism.

sCA tests have assumed timescales from 15 – 55 seconds. This range lies within the range of physiologic mechanism timescales, slower than myogenic and neurologic processes but faster (depending on the actual averaging chosen) than metabolic and endothelial processes. Due to mechanistic constraints discussed above, sCA testing will not result in metabolic or neurologic mechanism response action without endangering the patient. These tests may be further confounded by other regulatory mechanisms such as the compensatory reserve.

Based on these results, we can form hypotheses about the information different tests may be telling us. We hypothesize that dCA tests can have two types of results: one in which sCA tests also show impaired CA and one in which the two tests are in conflict. In the first type, vasoreactivity of all mechanisms may be impaired, therefore CBF compensation for the pressure drop will not occur. In this scenario, sCA tests will also reflect impaired CA. In the second type, the latency period of the myogenic mechanism is prolonged, or the mechanism is completely abolished, but other mechanisms remain intact. The baroreflex (or other autoregulatory reflexes) should then cause ABP to return to baseline before the CA can respond. Thus dCA tests would not observe CA function but sCA tests will.

### Informing correlation-based metrics

The choice of averaging window is essential in determining what CA mechanisms correlation-based metrics, such as PRx and Mx, are reflecting*(122)*. The evidence of any mechanism that acts on timescales faster than this window will either be minimized or averaged out. This is also the case if ABP is transiently changed faster than the averaging window. This averaging is purposeful, as it removes the fluctuation during the cardiac cycle, but could remove other important information. Consider the following example: if there is a transient change in ABP that is shorter than the width of the averaging window, then that change will be averaged out. Additionally, because the timescale of the myogenic mechanism (< 1 second) is shorter than most averaging window choices, the action of the myogenic mechanism in response to a transient ABP change will also be removed by averaging. In this scenario, the PRx or Mx measurement will be almost identical for impaired and non-impaired CA. Further, if the myogenic mechanism is impaired enough, such that this ABP perturbation results in reactive hyperemia (30 -120 seconds), then one would observe increased CBF and no change in ABP. The Mx calculation would be 0, which would falsely indicate *intactness* of the myogenic mechanism when it is actually impaired. Future work should be done to analyze how averaging windows affect PRx methodology for finding CPPopt and suggest methods to find the optimal averaging window based on the patient's status.

Additionally, if the correlation coefficient metrics do not account for time lags, correlation-based metrics may be strongly impacted by mechanistic time delays. For example, if the myogenic mechanism (action < 1 second) is impaired and the metabolic (action > 30 seconds) is not, CA will not occur for a significant amount of time after a large pressure change. Before this delay, $Mx \approx 1$, giving a pessimistic diagnosis when CA is partially intact. After this delay, the correlation would be $Mx \leq 0$ (given pressure is maintained), giving a far too optimistic diagnosis. This will also be the case if the location of blood pressure measurement is not accounted for. For example, if blood pressure is taken at an artery away from the head, there will be a delay between when the blood pressure reflects a perturbation and when that perturbation is seen in the CBF or ICP.

Finally, we address what mechanisms correlation-based metrics may be testing. A good PRx (or Mx) value ($\leq 0$) indicates that the CA mechanisms trigger the vasculature to change diameter in response to pressure gradient change rather than passively translating input pressure into intracranial pressure, possibly indicating that PRx is optimized in the center of the autoregulatory range. The metabolic mechanism acts independent of this range and can also shift this range. While PRx does not indicate metabolic mechanism functionality, it may be confounded by it. Therefore, optimal PRx may be found not due to CPP but due to an optimal metabolic environment. We recommend that existing indices of assessing autoregulation could be enhanced by other clinically related data.

## Outlook

This review presents timescale hypotheses on CA mechanisms, metrics, and tests. We propose that these hypotheses can be tested using computational models to evaluate the spatial and temporal scales on which assumptions may fail. Failure indicates that either: the mechanisms are interacting over that timescale, or the hypotheses are incorrect. Analysis of these mechanistic timescales will give great insight into the biological CA phenomenon and the complexity of its possible impairment.

 sCA metrics such as PRx and Mx are undoubtedly useful in the clinical setting. However, their interpretation should be considered very carefully and enhanced through the consideration of mechanisms beyond the myogenic mechanism. Based on this review, we suggest a simple workflow to check which CA mechanisms are intact. This workflow is not a clinical suggestion, but rather a simple thought experiment and first step in determining tests that isolate individual CA mechanisms.

dCA gives more direct insight into the impairment or extended latency of the myogenic mechanism than sCA. If dCA is impaired, the myogenic mechanism can be considered either impaired or highly delayed. One should observe the rate at which the blood pressure returns to base line. We suggest that the latency of the CA mechanism is longer than this time. The PRx window should be calculated according to this latency. A next step could be to induce a hypercapnic environment and repeat the test. If the environment does not influence PRx or dCA measurements, then the metabolic mechanism can be considered impaired, and steps should be made to improve the reactivity to metabolites. If dCA is seen to be intact, this environment should be considered when calculating CPPopt or trying to lower CBF. If dCA is still considered impaired, it can be concluded that the myogenic mechanism may be impaired (perhaps due to a cytosolic environment) and steps should be taken to improve vasoreactivity. Finally, hyperosmolar drugs can be used to indicate whether the endothelial mechanism is impaired. The decrease in blood viscosity will cause a decrease in shear stress, theoretically shifting the upper bound of the myogenic autoregulatory range to the right. If this range does not increase, the endothelial mechanism may be impaired.

Previous work demonstrates tremendous promise in CA-guided therapies on improving patient outcomes*(17–20)*. However, there are still significant obstacles involving metric incompatibility*(25)*. We suggest that some of this incompatibility is due to testing different underlying mechanisms, which leads to an exciting proposition: thus far, CA-guided therapy has been dominated by finding the pressure gradient (CPPopt) that optimizes CA functionality. Knowledge of the functionality of all CA sub-mechanisms may allow the autoregulatory range to be even further optimized. If we direct treatment toward not only the pressure gradient but the other CA mechanisms, patient outcomes could potentially be improved even further.

*Acknowledgements:*

We thank the Department of Bioengineering, College of Engineering, Design, and Computing, Aurora, CO, Section of Informatics and Data Science; Department of Pediatrics, University of Colorado School of Medicine, Aurora, CO; Section of Critical Care Medicine, Department of Pediatrics, University of Colorado School of Medicine, Aurora, CO; Departments of Neurology and Biomedical Informatics, Columbia University Vagelos College of Physicians and Surgeons, NewYork-Presbyterian Hospital; and Program for Hospital and Intensive Care Informatics, Department of Neurology, Columbia University. We also thank Dr. Richard Benninger for their helpful advice. This work was supported by NICHD R01HD105939 (to TDB), NSF grant DGE-1938058_Briggs (to JKB), grants from the National Institutes of Health RO1, "Mechanistic machine learning," LM012734, and "Discovering and applying knowledge in clinical databases" LM006910 " (to DJA), and R21NS113055 (to SP). The funders had no role in the study design, data collection, and analysis, decisions to publish, or preparation of the manuscript. The funders had no role in the design, decision to publish, or preparation of the manuscript.

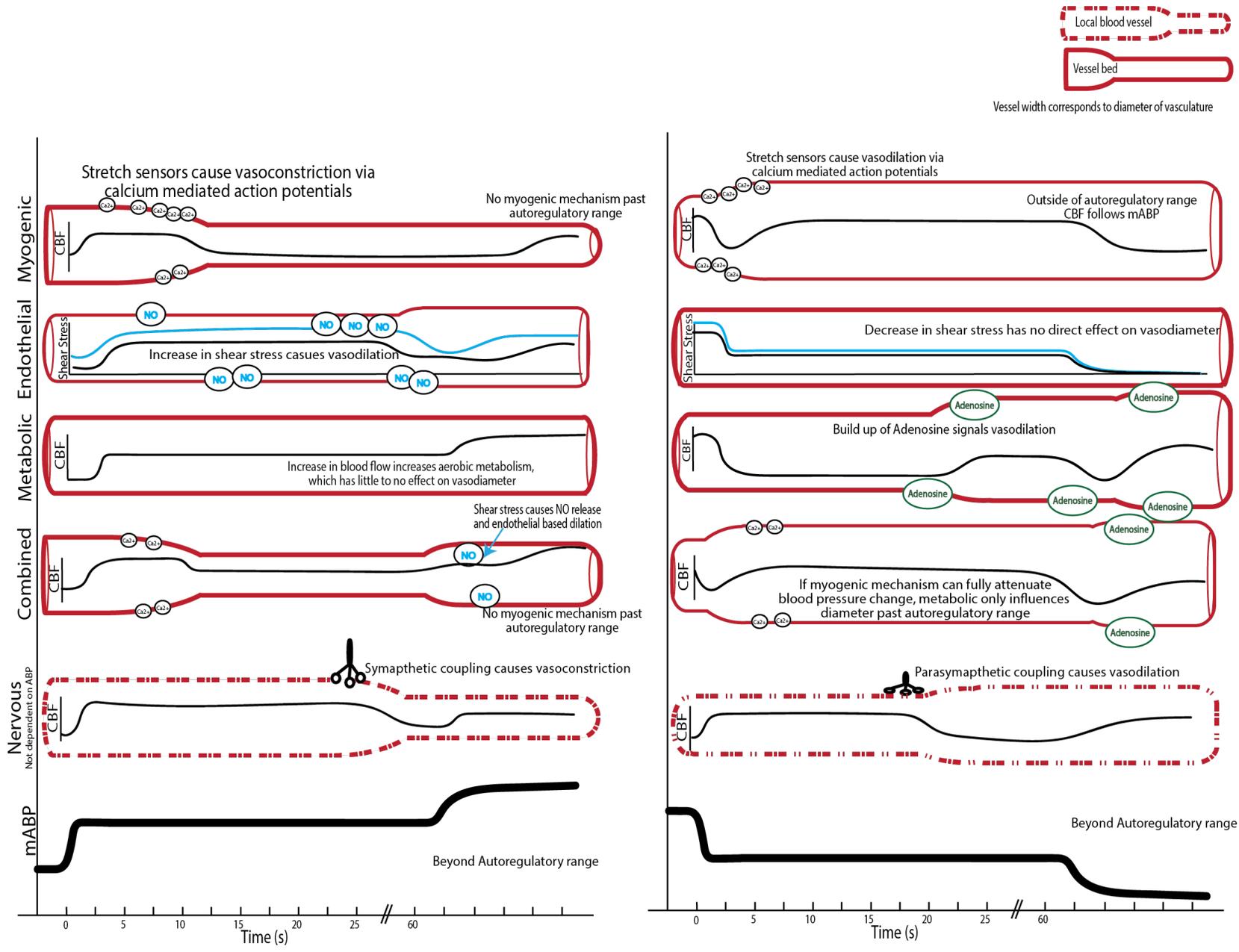

Figure 1)

**Table 1)**

| PROPOSED MECHANISM | | DETAILS | TIMESCALE | REACTION TO ABP (INDEPENDENT OF OTHER MECHANISMS) | PROPOSED INTERACTION WITH OTHER MECHANISMS |
|---|---|---|---|---|---|
| MYOGENIC | | Stretch sensitive ion channels initiate vasoconstriction or vasodilation upon pressure changes. | < 1s (49, 87) | Increase → Constriction<br>Decrease → Dilation | "First line of defense."(31) If mechanism can completely compensate for pressure change, no other mechanisms will react. |
| ENDOTHELIAL | | Shear stress on the endothelium causes the release of nitric oxide(10) and triggers vasodilation.<br>Shear stress is influenced by blood viscosity & blood flow<br>Prolonged shear stress causes vascular remodeling and possible dysfunction(123) | Onset: ~60 s (88)<br>Plateau: 1-2 min (88) | Increase → Dilation<br>Decrease → No direct change<br>Prolonged Increase → Vascular remodeling | Counteracts the effects of myogenic and metabolic<br>*Not prevalent in hypertensive animals* |
| METABOLIC | $CO_2$, PH, and lactate build up after pressure change | Accumulation of anaerobic metabolites ($pCO_2$ or pH) causes vasodilation, while excess $pO_2$ causes somewhat insignificant vasoconstriction(31)<br>Globally significant if CBF is reduced for a long enough time for the cerebrum to metabolize enough NADH and oxygen to resort to anaerobic metabolism (10).<br>Highly influential at localized, highly active brain regions. | 30 seconds of complete ischemia, multiple minutes of very low CPP (96)<br><br>Does not begin until CBF reaches $\sim \frac{25ml}{100g/min}$ (124) | Very large decrease in will cause anaerobic metabolism → Vasodilation<br><br>Increase → very minimal vasoconstriction | Globally: May act as a backup to myogenic mechanism according to slower timescale(31). Works outside of autoregulatory pressure ranges or if effects of myogenic mechanism has not fully compensated for pressure changes. |
| | Adenosine build up via pressure changes | Only metabolite shown to influence vasodiameter following changes in CPP in the autoregulatory range. | ~5 sec production(91)<br>Concentration dependent(93) | Decrease → Vasodilation<br><br>ABP increase → No effect | Locally: Works in conjunction with the neurologic mechanism to increase blood flow to regions of highly active brain(38). |
| | Clinically altered $CO_2$ | Hypoventilation will cause build up of $CO_2$ which will cause potent vasodilation.<br>Hyperventilation will cause removal of $CO_2$, which indirectly results in vasoconstriction. | Onset: hyperventilation ~20 sec (97, 99)<br>breath holding ~10 sec (98)<br>Plateau: ~ 30 sec | | $CO_2$ environment influences the myogenic mechanism's autoregulatory reserve: hypercapnia lowers upper reserve limit & hypocapnia increases lower reserve limit.(20) |
| NEUROLOGIC | | Neurotransmitters cause local vasodilation and vasoconstriction.<br>Occurs independently of pressure change. | Onset: .29 sec- 2 s (104, 105)<br>Plateau: 7 seconds | No direct reaction<br>*Parasympathetic and sympathetic nervous system activity can cause systemic blood pressure changes* | **Local:** Neuronal activity increases local metabolic need. |

**Figure 2)**

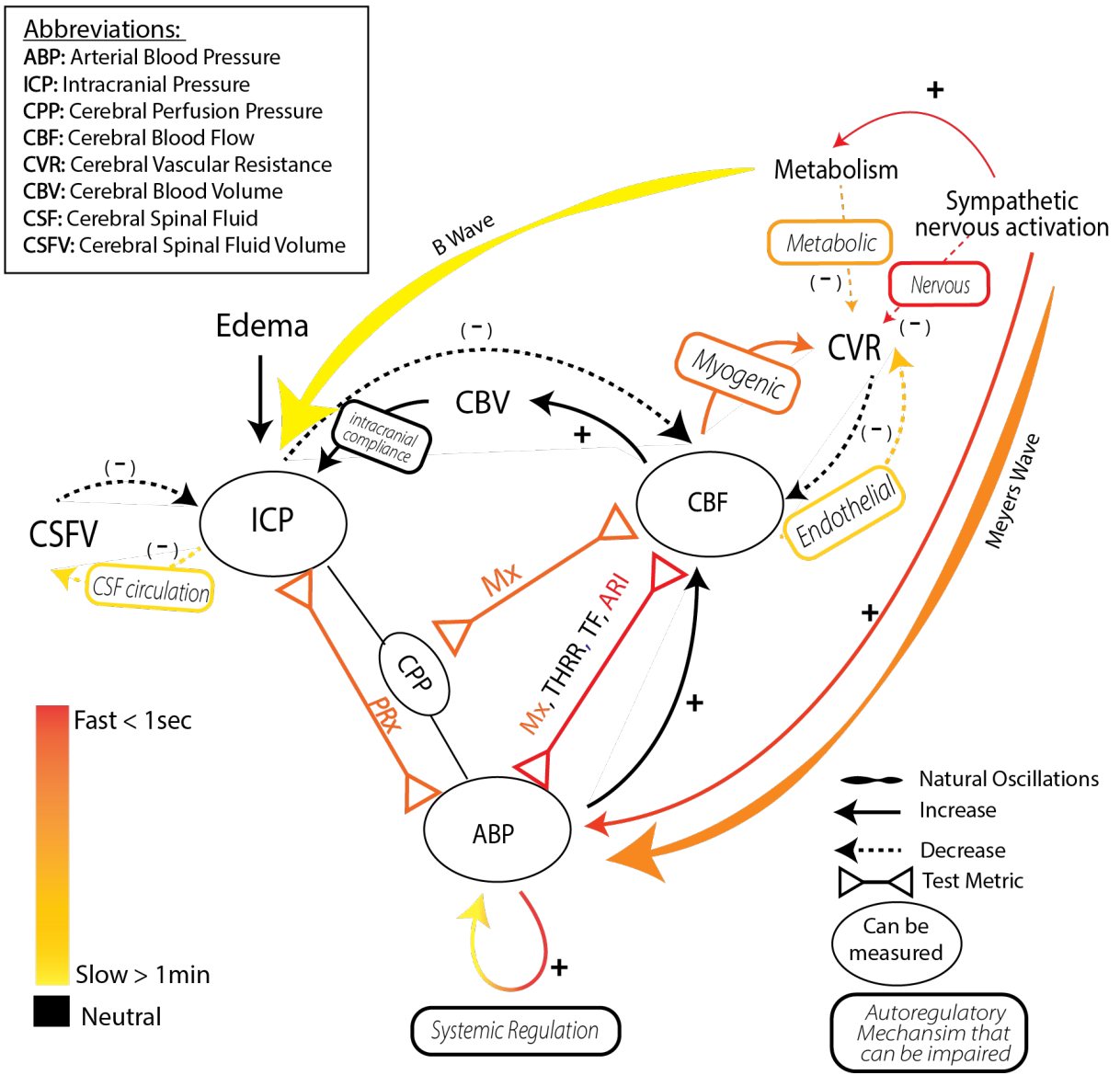

**Table 2**

| Dynamic or static CA | Metric | Metric Requirements | Assumptions | Timescale conclusions | Compatibility |
|---|---|---|---|---|---|
| sCA | Pressure Reactivity index (PRx) *(76, 78)* | Invasive ICP | Quasi-stationarity of mABP and ICP over averaging windows.<br><br>Only reflects timescales for amount of time data is being collected.<br><br>Relies on slow waves to measure correlation. | **15-55 sec***(107, 119)* | Shows a low correlation with Mx (R = 0.58*(125)* or .37*(25)*) |
| | Mean Flow index (Mx)*(48)* | Transcranial Doppler (TCD) | Sensitive averaging and correlation windows*(122)* | | |
| | TOxa*(126)* | Near Infrared spectroscopy (NIRS) | Quazi-stationarity over 10 seconds. | Tissue oxygenation shows coherence with CBF at 20-120 seconds therefore can be used in replacement for CBF in these frequencies *(126)* | Moderate correlation with Mx (R = .81*(126)* or .62*(25)*)<br>Low correlation with PRx (R = .39*(25)*) |
| | Transfer Function*(117)* | TCD | CA does not act on frequencies faster than heart rate.<br>Low signal coherence implies acting CA, further implying that without CA there is a linear relationship between pressure and flow. | **14-50** seconds*(117, 119)* | Non-significant linear relationship with ARI*(72)* |
| dCA | Transient hyperemic response ratio (THRR)*(74, 75)* | TCD<br><br>Carotid Compression | There will be dilation occurring during compression such that constriction can occur after release. | **1-2 sec** *latency*, plateau at **7 seconds** | |
| | Autoregulatory Index (ARI)*(68)* | TCD<br><br>Thigh Cuff | | **.65 sec** (ARI = 0) - **2 sec** (ARI=9) *latency* | |
| | Spectral analysis induced pressure change | TCD<br><br>Thigh cuff and rapid head-up tilt | | Onset: **2 s***(68, 116)*<br>Plateau: **5 sec** *(117)*-**7 sec** *(53)* | |
| Other | CO2 Vasoreactivity (Presented in table 1: Metabolic under Clinically altered CO2)*(127)* | Hyperventilation or inhaling CO2 | | **10-30 sec** *(97–99)* | |

# Figure Captions

**Figure 1:**

Depiction of how each of the four CA mechanisms is canonically hypothesized to respond to a rapid change in ABP (shown as a black graph bottom row). The left column demonstrates ABP increase and the right shows ABP decrease. Within each vessel (red) is a black graph indicating the response of CBF to the change in ABP and vascular diameter (characterized by vessel width). To the far right of each column, we show whether the mechanism is canonically proposed to function beyond the limits of the autoregulatory range. The top three rows show the theoretical response of the vessel, assuming all other mechanisms are absent (myogenic, endothelial, and metabolic, respectively). The fourth panel shows the combined effect of the three mechanisms above. Finally, the fifth row shows the response of the neurologic mechanism, which is independent of the ABP change.

**Figure 2:**

Depiction of the intracranial system timescales and complexity and how metrics attempt to capture system properties. The timescales are shown on a color gradient, from red (faster than 1 second), to yellow (slower than 1 minute). Clinically measurable quantities, including ABP, CBF, and ICP, are shown in black ovals. The four CA mechanisms are shown in colored rectangles with rounded edges. Also shown in rectangles are the systemic regulation, cerebral spinal fluid circulation, and intracranial compliance, which are other autoregulatory mechanisms that influence the cerebral hemodynamic system and can become impaired. Relationships between physiologic parameters are shown as dashed lines from *A* to *B* if *A* causes *B* to decrease or solid lines if *A* causes *B* to increase. CA metrics are drawn as inverted arrows between the measurements which they use as inputs. Finally, metrics and their timescale assumptions are presented as inverted arrows connecting the quantities they use as inputs.